\documentclass{optica-article}
\journal{opticajournal} 
\articletype{Research Article}

\usepackage{lineno}
\usepackage{comment}

\begin{document}

\title{Saturable absorption in NV-doped diamond studied by femtosecond Z-scan}

\author{Wojciech Talik, Mariusz Mrózek, Adam M. Wojciechowski and Krzysztof Dzierżęga\authormark{*}}

\address{Marian Smoluchowski Institute of Physics, Jagiellonian University, ul. Łojasiewicza 11, 30-348 Kraków,  Poland}

\email{\authormark{*}krzysztof.dzierzega@uj.edu.pl} 

\noindent{\small\today}


\begin{abstract*} 
We investigate nonlinear optical absorption in diamond crystals containing high densities of nitrogen–vacancy (NV) centers using open-aperture Z-scan measurements with 230-fs laser pulses at 1032 nm, within the transparency window of diamond. While high-purity electronic-grade diamond exhibits third-order nonlinear absorption, NV-doped samples display pronounced saturable absorption that strengthens with increasing defect concentration. Linear transmission spectroscopy reveals that, in addition to NV centers, the crystals host significant populations of H2 (NVN$^{-}$) defect complexes whose absorption band partially overlaps the excitation wavelength. By correlating spectroscopic data with nonlinear measurements and modeling the response using an effective two-level system, we show that the observed saturation cannot be attributed solely to NV centers but arises from the combined contribution of NV-related and H2 defects. For the highly doped sample, we determine an effective linear absorption coefficient of $\alpha_0 \approx 6.52~\text{cm}^{-1}$ and a saturation intensity of $I_s \approx 40.0~\text{GW/cm}^2$. These findings highlight the critical role of the complex defect landscape in governing the nonlinear optical response of NV-doped diamond and underscore the necessity of accounting for ancillary defect species in the design of diamond-based nonlinear and quantum photonic devices.


\end{abstract*}

\section{Introduction}

In recent decades, diamond has emerged as one of the most promising materials for applications in quantum optics and nanotechnology. This interest is driven by its exceptional physical properties, including high thermal conductivity, extreme mechanical hardness, chemical inertness, and a broad optical transparency window. Together, these characteristics make diamond an ideal host matrix for point defects exhibiting remarkable quantum properties. Among these defects, nitrogen-vacancy (NV) color centers play a central role. Found predominantly in the negative charge state (NV$^{-}$)\cite{Doherty_2013}, these centers combine long spin coherence times with efficient initialization and readout, enabling a wide range of applications in quantum sensing, communication, and information processing \cite{Hall_2013, Rizzato_2023, Xu_2019}.

The fabrication of diamond samples containing high-density NV ensembles typically involves nitrogen doping, followed by electron-beam irradiation and subsequent thermal annealing. While this approach effectively increases the NV concentration, it also unavoidably leads to the formation of additional lattice defects \cite{Torelli_2020}. As a result, NV centers coexist with other defect species, such as substitutional nitrogen (N$_{s}^{0}$) and more complex aggregates, including H2 (NVN$^{-}$) and H3 (NVN$^{0}$) centers \cite{Rand_1985, Wong_2022, Buerki_1999}. 

To fully exploit diamond-based quantum technologies, it is therefore essential to move beyond the linear optical regime and develop a detailed understanding of how these defect-rich crystals interact with intense optical fields. Femtosecond laser pulses provide a powerful means to probe such interactions, enabling the study of nonlinear optical phenomena such as multiphoton absorption and the optical Kerr effect (OKE) while minimizing thermal damage to the crystal lattice. These nonlinear processes underpin a variety of emerging applications, including ultrafast control of NV-center spin states, laser-assisted creation and manipulation of defect centers, and the realization of diamond-based integrated photonic platforms \cite{Sato_2025}. Moreover, insight into the ultrafast optical dynamics of defect centers is crucial for the advancement of quantum sensing protocols, which often rely on high-power optical excitation to enhance sensitivity.

Despite the extensive body of work on NV centers for metrology and magnetic field sensing, their nonlinear optical (NLO) response remains comparatively underexplored. Early investigations include four-wave mixing experiments by Rand \cite{Rand_1988} and pioneering studies by Sheik-Bahae \textit{et al}. \cite{Sheikbahae_1995}. More recently, Motojima \textit{et al.} \cite{Motojima_2019} employed the Z-scan technique to demonstrate the influence of near-surface NV centers on the optical Kerr effect and two-photon absorption under femtosecond excitation at 800~nm. In contrast, Abulikemu \textit{et al.} \cite{Abulikemu_2021, Abulikemu_2022} reported second-harmonic generation and cascaded third-harmonic generation in bulk NV-doped diamond using longer-wavelength excitation in the 1200--1600~nm range. While nonlinear absorption effects have been observed in several studies, a comprehensive understanding of how individual defect species contribute to the overall nonlinear response -- particularly in diamonds with high color-center densities -- remains incomplete.\\

In our previous work \cite{Talik_2025}, we addressed this issue by investigating the nonlinear refractive index ($n_2$) of NV-doped diamond crystals via OKE measured using the Z-scan technique with 230-fs laser pulses at a wavelength of 1032~nm. This wavelength is red-detuned from the principal NV optical transitions and is approximately twice the wavelength typically used for NV excitation. We found that all studied samples, both undoped and NV-doped, exhibit a positive $n_2$ coefficient with a pronounced four-fold rotational anisotropy. Notably, NV-doped diamonds displayed lower $n_2$ values than pure diamond. This reduction was then attributed to a negative contribution of NV centers to the third-order susceptibility, modeled as a two-level system driven by a strong laser field detuned from its effective resonance. At the low repetition rate of 1~kHz used in the experiments, the 1042~nm singlet-singlet transition of NV$^-$ centers can be neglected, as the corresponding states remain unpopulated.

Building on these findings, the present work focuses on the nonlinear absorption properties of the same set of diamond crystals. We demonstrate that, at a wavelength of 1032~nm, the samples exhibit clear signatures of saturable absorption (SA), despite the excitation being far red-detuned from the main NV resonance. By correlating Z-scan measurements with a detailed analysis of linear transmission spectra, we show that this nonlinear response cannot be attributed solely to NV centers. Instead, a substantial contribution arises from H2 defects, whose absorption band partially overlaps with the excitation wavelength. These results highlight the crucial role of complex defect landscape in determining the nonlinear optical behavior of NV-doped diamond and underscore the importance of disentangling the contributions of individual defect species for the rational design of diamond-based nonlinear and quantum photonic devices.

\section{Experiment}
We investigated three commercially available diamond samples with distinctly different defect compositions. These included a high-purity electronic-grade single crystal (EGSC) and two nitrogen-vacancy-doped diamonds with medium and high NV concentrations of 0.3~ ppm (MCNV) and 4.5~ppm (HCNV), respectively. The latter two samples were fabricated using a standard process comprising nitrogen doping, electron irradiation, and subsequent thermal annealing. As a consequence, these crystals are expected to contain a high density of various point defects in addition to NV centers. 
The EGSC crystal was supplied by Element Six, while the MCNV and HCNV samples were obtained from Thorlabs. The key properties of all investigated samples are summarized in Table \ref{tab:sample-characteristics}. 

\subsection*{Transmission measurements}

Linear transmission spectra were recorded in the 200-1100~nm spectral range using a dual-beam UV/VIS/NIR spectrophotometer (SPECORD210 Plus, Analytik Jena) with a spectral bandwidth of 1.0~nm. During the measurements, the crystals were mounted in custom-designed 3D-printed holders to ensure reproducible alignment and stable positioning.\\
\begin{table}[!hbt]
\centering
\small
\caption{\textbf{Characteristics of the investigated diamond crystals.}}
\label{tab:sample-characteristics}
\begin{tabular}{ccccccc}
\hline
Designation & N [ppm] & NV [ppm] & Size [$\rm mm^3$]  & Face & Edges  & Supplier \\
\hline
EGSC & $< 0.005$ & $-$ & $2 \times 2 \times 0.5$ & $(001)$ & $[110]$ & Element Six\\
MCNV & $0.8$ & $0.3$ & $3\times3\times0.5$ & $(001)$ & $[100]$ & Thorlabs / Element Six\\
HCNV & $13$ & $4.5$ & $3\times3\times0.5$ & $(001)$ & $[100]$ & Thorlabs / Element Six\\
\hline
\end{tabular}
\end{table}
All samples exhibit a sharp decrease in transmission below approximately 250~nm, corresponding to the onset of intrinsic valence-to-conduction band transitions in diamond. Apart from this absorption edge, the transmission spectrum of the undoped EGSC crystal is featureless across the UV-NIR range and is primarily limited by Fresnel reflection losses at the crystal surfaces. In contrast, both the MCNV and HCNV samples show pronounced and broadband absorption features throughout the measured spectral range, consistent with the presence of multiple optically active defects \cite{Talik_2025}.

\begin{figure}[!bth]
  \centering
  \includegraphics[width = 0.95\textwidth]{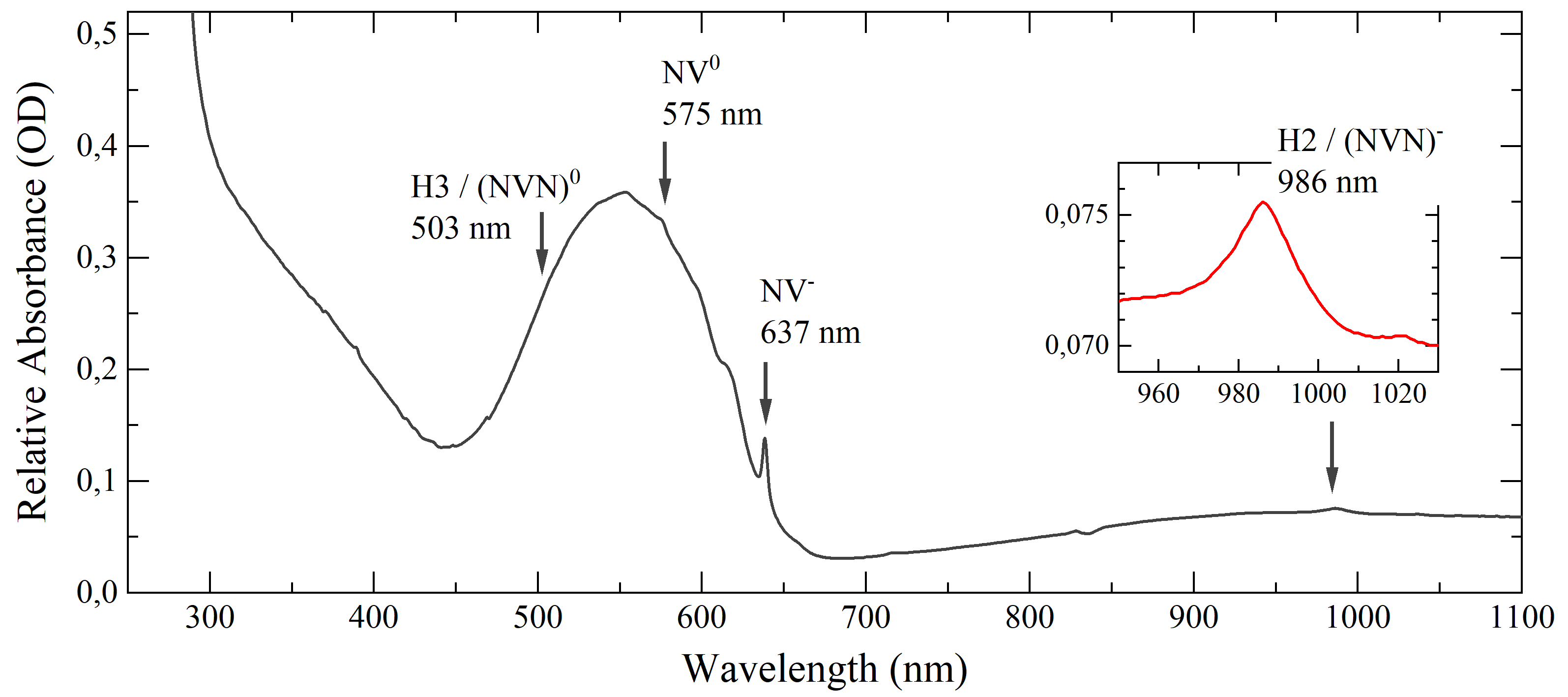}
  \caption{Relative absorbance spectra of the HCNV crystal measured with MCNV crystal in the reference channel of the spectrometer with marked zero-phonon lines (ZPL) of NV, H2 and H3 point defects. In the inset ZPL of $\rm NVN^-$ center at 986~nm.  The feature near 840~nm is an artefact caused by spectrometer malfunction in this spectral region.
 }
  \label{fig:transmission_spectra}
\end{figure}

To identify the origin of these features, differential absorbance measurements were performed by placing the HCNV and MCNV crystals in the signal and reference arms of the spectrometer, respectively. This configuration suppresses geometric effects and Fresnel reflection contributions. The resulting relative absorbance spectrum is shown in Fig.~\ref{fig:transmission_spectra} and reveals three dominant absorbance bands: in the near-infrared (750-1100~nm), visible (460-650~nm), and ultraviolet (below 450~nm) regions.
These bands can be attributed to electronic and vibronic transitions associated with NV-related and N-V-N defect complexes. Distinct zero-phonon lines (ZPLs) are clearly identifiable at 986~nm (H2), 637~nm ($\rm NV^-$), 575~nm ($\rm NV^0$), 503~nm (H3), and 496~nm (H4) \cite{Buerki_1999, Seo_2011, Johnson_2025}. This spectral signature confirms that, in addition to nitrogen-vacancy centers, the crystals also contain N-V-N complexes in both negative (H2) and neutral (H3) charge states, as well as the more complex N-V-N-V-N aggregate (H4).
Such defect complexes are known to form in diamonds subjected to high nitrogen concentration followed by electron irradiation and annealing at temperatures between 600-1200~$\rm ^o C$ \cite{Torelli_2020}. 
The pronounce absorption below 450~nm is attributed to overlapping contributions from the vibronic sidebands of H3 and H4 centers, photoionization transitions of substitutional nitrogen ($\rm N_s^0$) and $\rm NV^-$ centers, as well as additional irradiation-induced defect complexes. Broadband UV absorption of this type is a well established characteristic of nitrogen-rich, irradiated diamonds and is commonly associated with defect-related electronic states located deep in the bandgap \cite{Kazuchits_2016}.

\subsection*{Open-aperture Z-scan measurements}

\begin{figure}[b!]
    \centering
    \includegraphics[width=0.75\linewidth]{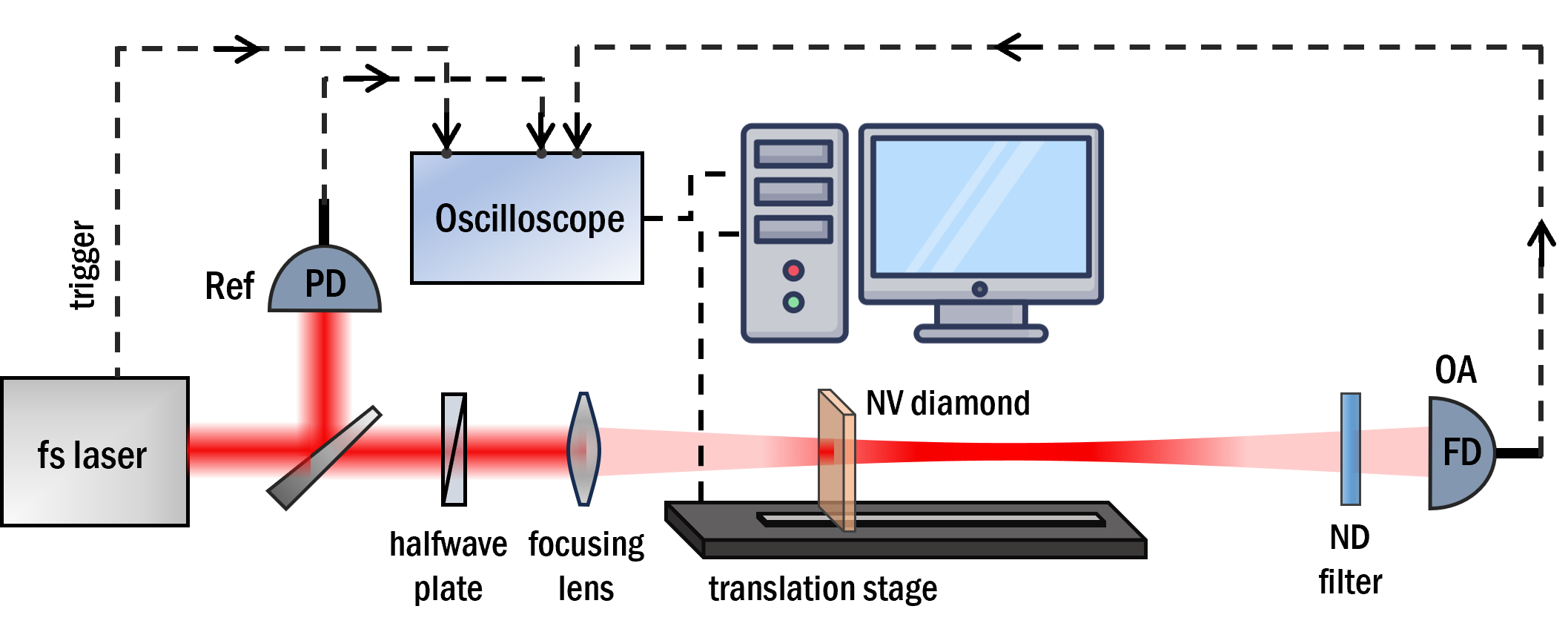}
    \caption{Schematic of the open-aperture (OA) Z-scan setup used to study nonlinear absorption.}
    \label{Fig: ExpSetup}
\end{figure}

Nonlinear optical absorption in the diamond crystals was investigated in the infrared transparency window at a wavelength of 1032~nm using the Z-scan technique \cite{Sheikbahae_1990a}. All measurements were performed in the open-aperture (OA) configuration, which is sensitive exclusively to intensity-dependent absorption processes. The experimental arrangement followed the standard Z-scan geometry and is schematically illustrated in Fig.~\ref{Fig: ExpSetup}. 

Ultrashort laser pulses were generated by an Yb-fiber laser system (Jasper Flex, Fluence) delivering 230~fs pulses at a central wavelength of 1032~nm (photon energy 1.201~eV). The laser provides a maximum repetition rate of 10~MHz and a maximum pulse energy of 10~$\mu$J. For the present experiments, the repetition rate was deliberately reduced to 1~kHz in order to suppress potential cumulative thermal effects and ensure that the measured nonlinear response was purely electronic in origin. 
A fraction of the incident beam was diverted by a beam splitter and directed to a reference detector for continous monitoring of pulse-to-pulse energy fluctuations. The transmitted beam was focused using a lens with a focal length of $f=10$~cm, resulting in a focal spot diameter of $2w_0=70$~$\mu$m and a Rayleigh range of $2z_0=7.3$~mm. 
During the Z-scan measurement, the sample was translated along the beam propagation axis ($z$) through the focal region, thereby varying the peak intensity experienced by the crystal.

Both the OA and reference signals were detected using identical large-area photodiodes (DET100A/M, Thorlabs) and recorded with a high-bandwidth digital oscilloscope (1~GHz, 10~GS/s, Tektronix). The pulse energies were extracted by integrating the temporal waveforms of the detected pulses. For each laser shot, the OA was normalized to the simultaneously recorded reference signal to eliminate the laser intensity fluctuations. 
At each sample position, the normalized values were averaged over 500 consecutive laser pulses before being transferred to a computer for further analysis. This averaging procedure significantly enhanced the signal-to-noise ratio and ensured high measurement reproducibility. Preliminary damage-threshold test confirmed that none of the investigated crystals exhibited permanent optical or structural modifications for peak intensities below 500~GW/cm$^2$, corresponding to pulse energies of approximately 2~$\rm \mu J$. All Z-scan measurements reported in this work were therefore conducted below this threshold.

\section{Results and discussion}

Figures \ref{fig:OA_EGSC_MCNV} and \ref{fig:OA_HCNV} present the normalized transmittance obtained from OA Z-scan measurements for the EGSC, MCNV, and HCNV diamond crystals at various on-axis peak intensities \( I_0 \) at the focal plane. A clear qualitative difference in nonlinear optical response is observed between the undoped and NV-doped samples. While the EGSC crystal exhibits \textbf{nonlinear absorption} (NA), both NV-doped crystals display \textbf{saturable absorption} (SA), which becomes increasingly pronounced with increasing nitrogen and NV concentrations and is stronger in the HCNV sample. 
The solid curves in the figures represent theoretical fits of the normalized OA transmittance \( T_{\rm OA}(z) \) to the experimental data as a function of the sample position $z$.

\subsection*{Nonlinear absorption in EGSC crystal}

For the EGSC crystal, the nonlinear absorption is described by an intensity-dependent absorption coefficient of the form
\begin{equation*}
   \alpha(z)=\alpha_0 + \beta I(z), 
\end{equation*}
where $\alpha_0$ is the linear absorption coefficient and $\beta$ is the nonlinear absorption coefficient. 

\begin{figure}[!b]
  \centering
  \includegraphics[width = 0.95\textwidth]{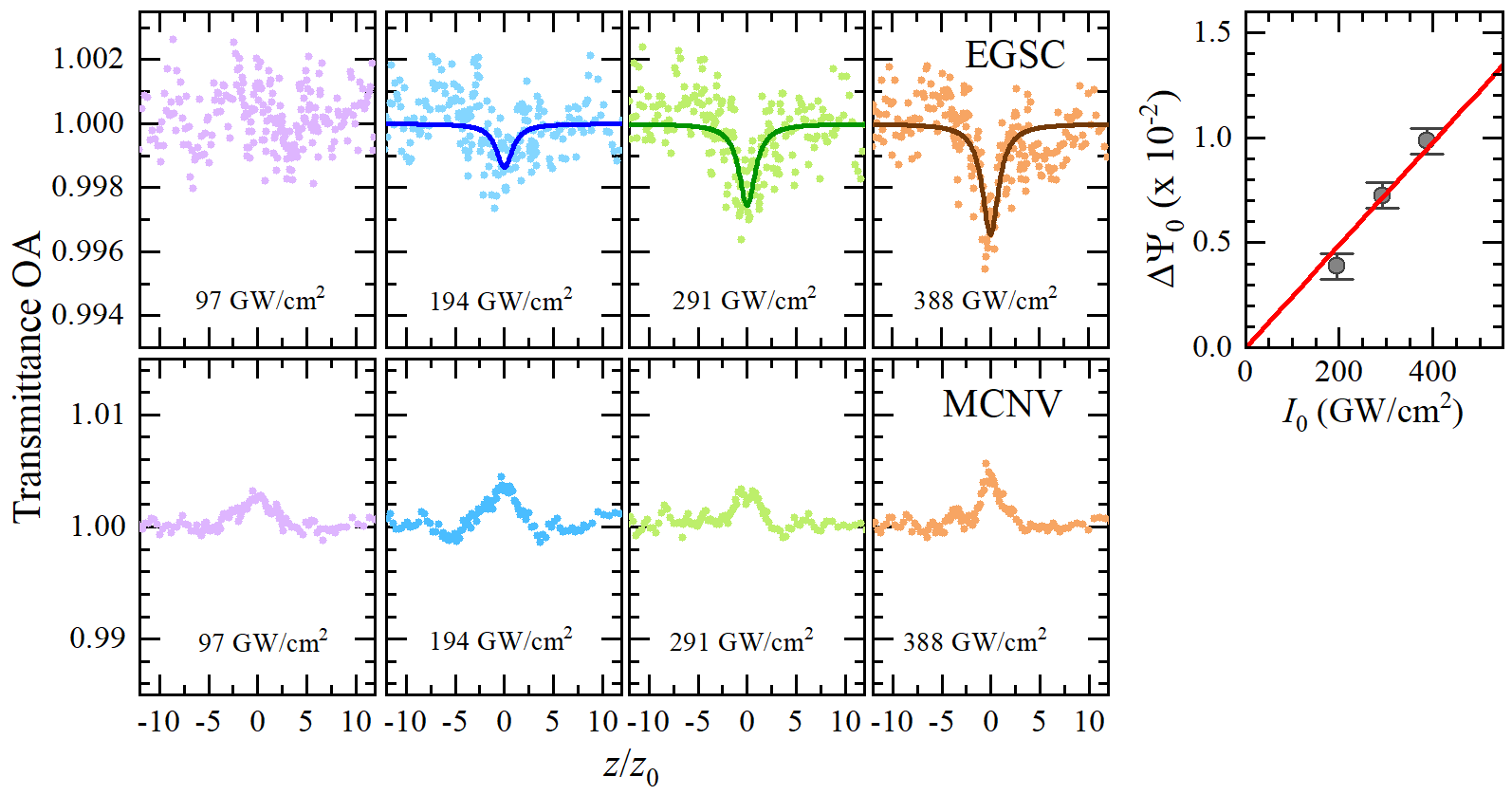}
  
  \caption[Results of OA Z-scan measurements for EGSC diamond.]{Normalized transmittance measured by the open-aperture (OA) Z-scan method for pure diamond, \textbf{EGSC} (upper row), and low concentration NV-doped diamond, \textbf{MCNV} (lower row), at various laser beam intensities $I_0$. The solid lines represent the fits of Eq.~(\ref{eq:ZScanOA_NAbs}) to the experimental data. The upper-right figure shows the nonlinearity parameter $\Delta \Psi_0$ as a function of the laser beam intensity $I_0$. The determined nonlinear absorption coefficient at $\lambda=1032$~nm is $\beta=4.9(3) \times 10^{-15}$~m/W.}
  \label{fig:OA_EGSC_MCNV}
\end{figure}

The position-dependent laser intensity is given by $I(z)=I_0 /(1+z^2/z_0^2)$, where $z_0$ is the Rayleigh range. Assuming a Gaussian spatial beam profile and a Gaussian temporal pulse envelope, the normalized OA transmittance can be expressed as \cite{Sheikbahae_1990a} 
\begin{equation}
    T_{\rm OA}^{\rm NA}(z) = \sum_{m=0}^{\infty} \frac{[-\Delta\Psi_0(z)]^m}{(m+1)^{3/2}(1+z^2/z_0^2)^m}, 
\label{eq:ZScanOA_NAbs}
\end{equation}
where $\Delta\Psi_0=\beta\, I_0\ L_{\rm eff}$, $L_{\rm eff}=[1-\exp{(-\alpha_0\,L})]/\alpha_0$ and $L$ is the crystal thickness. 
In our case, truncation of the series in Eq.~(\ref{eq:ZScanOA_NAbs}) at \( m = 3 \) yields sufficient numerical accuracy.

The OA Z-scan data for the EGSC crystal are shown in Fig.~\ref{fig:OA_EGSC_MCNV} (upper row), together with fits of Eq.~(\ref{eq:ZScanOA_NAbs}). The extracted nonlinearity parameter $\Delta\Psi_0$ increases linearly with $I_0$, confirming that the observed response is dominated by second-order nonlinear absorption. From the slope of this dependence, we obtain a nonlinear absorption coefficient of $\beta=4.9(3) \times 10^{-15}$~m/W at wavelength of 1032~nm.

Given the large indirect band gap of diamond ($\approx5.5$~eV), direct two-photon excitation across the band gap is unlikely at this photon energy. The observed nonlinear absorption is therefore attributed to two-photon excitation of deep impurity-related states or residual lattice defects, which are known to be present even in high-purity electronic-grade diamond.

\subsection*{Saturable absorption in NV-doped crystals}

In contrast to the EGSC sample, both NV-doped crystals (MCNV and HCNV) exhibit saturable absorption. In this case, the absorption coefficient is modeled as
\begin{equation}
    \alpha(z)=\frac{\alpha_0}{1+I(z)/I_s}
    \label{eq:nonl refraction_SatNAbs}
\end{equation}
where $I_s$ is the saturation intensity. This expression can be rewritten as
\begin{equation}
    \alpha(z)= \alpha_0-\alpha_0\,\frac{I(z)/I_s}{1+I(z)/I_s},
    \label{eq:nonl refraction_SatNAbs}
\end{equation}
highlighting the intensity-dependent reduction of absorption.
The corresponding normalized OA transmittance is given by \cite{Sheikbahae_1990a}

\begin{equation}
    T_{\rm OA}^{\rm SA}(z) = \sum_{m=0}^{\infty} \frac{[\alpha_0\,L_{\rm eff})]^m}{(m+1)^{3/2}} \left( \frac{I_0/I_s}{1+I_0/I_s+z^2/z_0^2} \right)^m. 
    \label{eq:ZScanOA_SatAb}
\end{equation}

\begin{figure}[!bth]
  \centering
   \includegraphics[width = 0.95\textwidth]{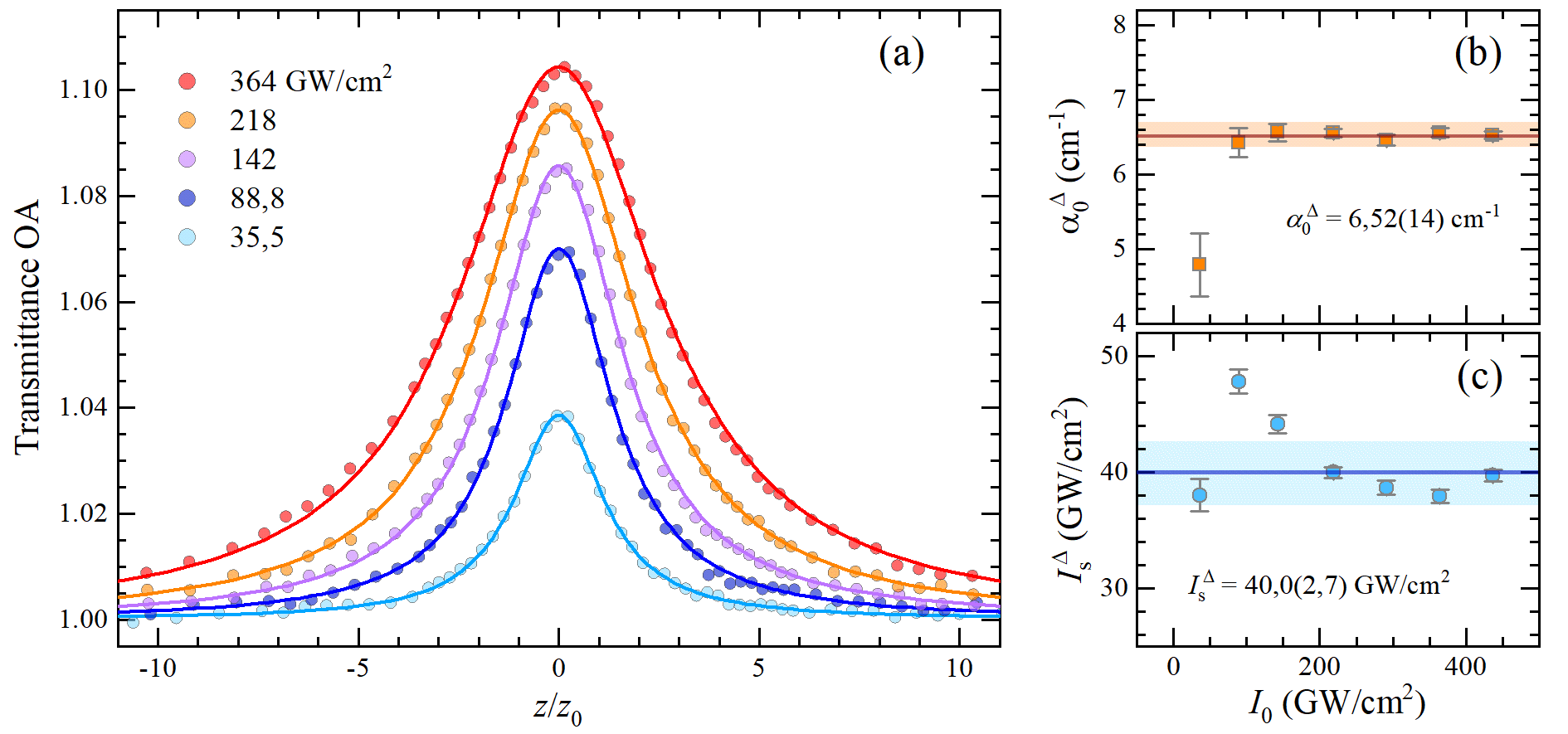}
   \caption{(a) -- Normalized transmittance measured by the open-aperture (OA) Z-scan method for diamond with a high concentration of NV color centers (\textbf{HCNV}), for various laser beam intensities $I_0$. The solid lines represent the fit of Eq.~(\ref{eq:ZScanOA_SatAb}) to the experimental data. (b,\,c) –- dependence of the linear absorption coefficient $\alpha_0$ and the saturation intensity $I_s$, respectively, obtained from the fit, on the intensity $I_0$.}
  \label{fig:OA_HCNV}
\end{figure}

Figure~\ref{fig:OA_HCNV}(a) shows the OA Z-scan traces for the HCNV crystal along with fits of Eq.~(\ref{eq:ZScanOA_SatAb}). The extracted fitting parameters -- the linear absorption coefficient $\alpha_0$ and saturation intensity $I_s$ -- are summarized in Figs.~\ref{fig:OA_HCNV}(b, c). Both parameters remain independent of the applied peak intensity $I_0$ Within the statistical uncertainty, with minor deviations recorded for lower intensities only, confirming the internal consistency of the SA model. The weighted mean values obtained for the HCNV crystal are $\alpha_0=6.52(14)$ cm$^{-1}$ and $I_s=40.0(2.7)$ GW/cm$^2$.
The pronounced saturable absorption observed in NV-doped samples, and its absence in the undoped EGSC crystal, indicate that this nonlinear response originates from defect-related electronic transitions and not from the diamond lattice itself.

\subsection*{Two-level model for saturable absorption}
In order to identify the microscopic origin of the SA effect and to determine which defect species contribute most strongly to the observed absorption saturation, we analyze the experimental results within the framework of a two-level system model.

\begin{figure}[!bht]
  \centering
  \includegraphics[width = 0.95\textwidth]{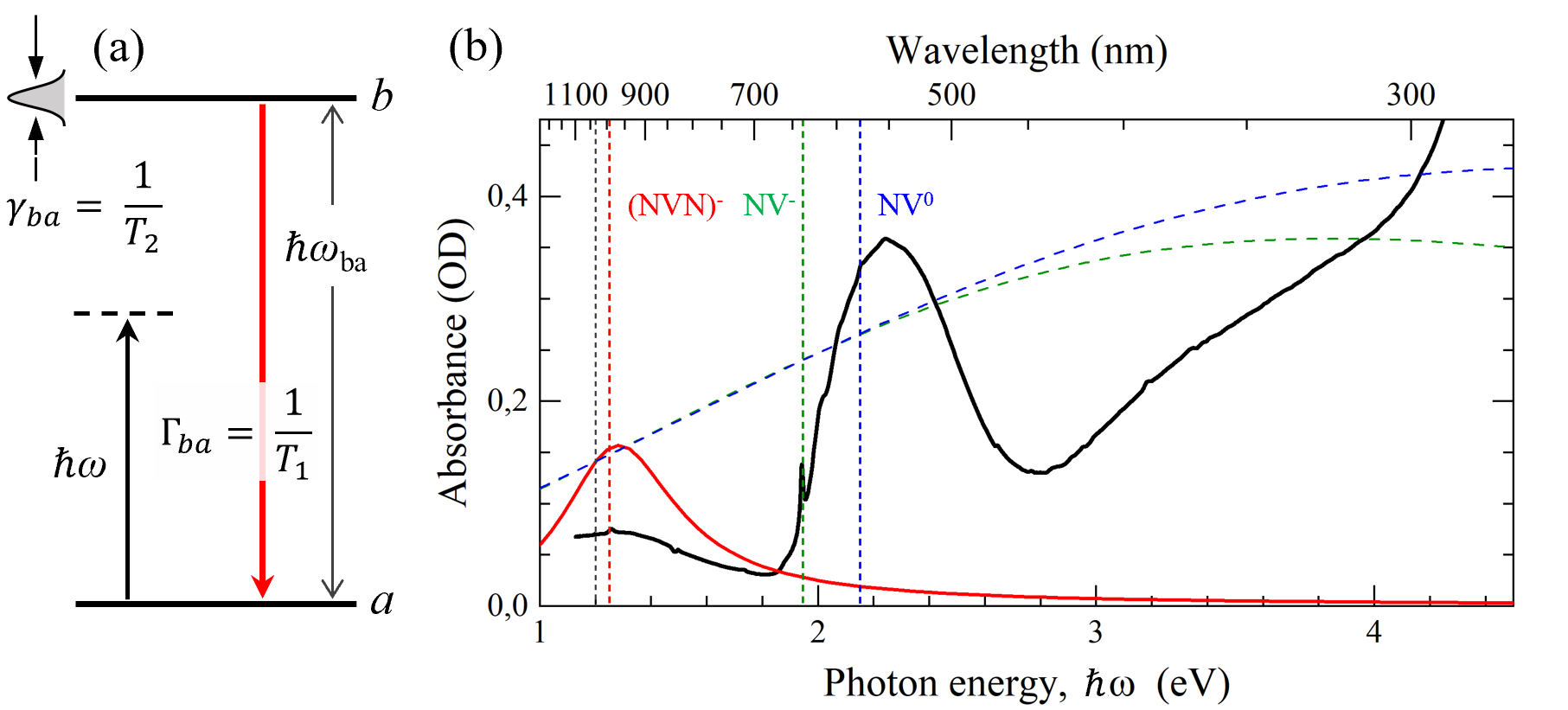}
  \caption{(a) Schematic representation of point-defect centers modeled as effective two-level systems with resonance frequency $\omega_{\mathrm{ba}}$, interacting with a nonresonant laser field of photon energy $\hbar\omega$. The parameters $T_1$ and $T_2$ denote the effective population relaxation time of the excited state \(b\) and the coherence decay (dephasing) time of the induced dipole moment, respectively. (b) Absorbance spectra measured using the spectrophotometer (solid black) and reconstructed using Eq.~(\ref{eq: Isat od Delta}) (solid red), with the coherence decay time $T_2$ inferred from combined closed-aperture (CA) and open-aperture (OA) Z-scan measurements. The vertical black dashed line indicates the excitation laser frequency, while the colored dashed lines denote the resonance frequencies of the considered defect centers.}
  \label{fig:NV_2level_model}
\end{figure}

Saturable absorption is most commonly described using a simple two-level system interacting with a strong optical field. Within this approach, each optically active point-defect in NV-doped diamond can be approximated as an effective two-level system, as schematically illustrated in Fig.~\ref{fig:NV_2level_model}(a). 
The ground and excited states, denoted by \(a\) and \(b\), represent effective electronic states that incorporate the associated phonon sidebands. This approximation is well justified by the broad absorption features observed in the linear transmission spectra shown in  Fig.~\ref{fig:transmission_spectra}.

Following the treatment of Boyd \cite{Boyd_2020}, the optical susceptibility of a two-level system driven by a near-resonant electromagnetic field is given by
\begin{equation}
    \chi=-N |\mu_{\mathrm{ba}}|^2 \frac{T_2}{\epsilon_0 \hbar} \frac{\Delta T_2 - i}{1+\Delta^2 T_2^2 +|E|^2/|E_s^0|^2},
\end{equation}
where $\Delta=\omega-\omega_{\mathrm{ba}}$ is the detuning of the driving field $\omega$ from the transition frequency $\omega_{\mathrm{ba}}$, $\mu_{\mathrm{ba}}$ is the transition dipole moment, and $N$ is the concentration of optically active defects. The parameters $T_1$ and $T_2$ denote the population relaxation time and the coherence decay time, respectively. The characteristic saturation field strength is defined as $|E_s^0|^2=\hbar^2/(4 |\mu_{\mathrm{ba}}|^2 T_1 T_2)$.

In the limit of weak optical susceptibility ($\chi \ll 1$), the intensity-dependent absorption coefficient can be approximated as
\begin{equation}
    \alpha(I, \Delta)\approxeq \frac{\omega}{c}\ \Im (\chi)=\left( 1+\frac{\Delta}
    {\omega_{\rm ba}} \right) \frac{\omega_{\rm ba}}{c}\ \Im (\chi)
    =\alpha_0(\Delta)\, \frac{1}{1+I/I_s(\Delta)},
    \label{eq:NLAbs_definition}
\end{equation}
where the optical intensity $I=\tfrac{1}{2}\varepsilon_0\, c\, n_0 |E|^2$. 
The experimentally determined linear absorption coefficient and saturation intensity therefore correspond to the detuning-dependent expressions
\begin{eqnarray}
    \alpha_0(\Delta) &=& \alpha_0\ (1+\Delta^2\,T_2^2)^{-1} \left( 1+ \frac{\Delta}{\omega_{\rm ba}}\right) \nonumber\\
    I_s(\Delta) &=& I_s^0\,(1+\Delta^2\, T_2^2)
    \label{eq: Isat od Delta}
\end{eqnarray}
where the resonant ($\Delta=0$) values are given by
\begin{eqnarray}
    \alpha_0 &=& \frac{\omega_{ba}}{\varepsilon_0\, c\, \hbar}\ N |\mu_{\rm ba}|^2\, T_2 \nonumber\, ,\\
    I_s^0 &=& \frac{\varepsilon_0\, c\, \hbar^2}{8 |\mu_{\rm ba}|^2}\ \frac{n_0}{T_1\, T_2},
    \label{eq:Isat rezonansowe}
\end{eqnarray}
Here, $n_0$ = 2.3904 is the refractive index of diamond at the excitation wavelength of 1032~nm \cite{Polyanskiy_2024}. 
Moreover, as shown by Talik \textit{\textit{et al.}} \cite{Talik_2025}, the nonlinear refractive index associated with the same two-level system is given by
\begin{equation}
    n_2 (\Delta) = N |\mu_{\rm ba}|^4 \frac{T_1 T_2^2}{n_0^2\, \varepsilon_0^2\, c \hbar^3}\frac{\Delta T_2}{(1+\Delta^2 T_2^2)^2}\,.
    \label{eq:n2Model_n2}
\end{equation}
Combining the above expressions yields a direct relation between the coherence decay time $T_2$ and experimentally accessible parameters obtained from the Z-scan measurements, namely $\alpha_0(\Delta)$, $I_s(\Delta)$, and $n_2 (\Delta)$ 
\begin{equation}
    T_2 = \frac{8\,n_0\,n_2}{c}  \frac{I_s(\Delta)}{\alpha_0(\Delta)}\ \frac{\omega}{\Delta}.
    \label{eq:T2}
\end{equation}
Here, according to Talik \textit{et al.}, we assume $n_{\rm 2}=-0.758(56) \times 10^{-20}$ m$^2$/W  as the contribution of point-defects to the overall nonlinear absorption coefficient in HCNV diamond crystal.
As one can see the only unknown in the above equation is the detuning $\Delta$, i.e. the resonance frequency $\omega_{\rm ba}$ of the considered two-level system.
Assuming that the observed saturable absorption effect originates from a specific (individual) point-defect, one can determine the effective coherence decay time $T_2$ and then, using Eq.~(\ref{eq: Isat od Delta}), reconstruct the spectrum of the absorption coefficient associated with this center.

Figure~\ref{fig:NV_2level_model}(b) shows the absorbance spectrum reconstructed for the HCNV crystal (red line), assuming that SA originates solely from H2 (NVN$^-$) defects with a resonance energy $\hbar\,\omega_{\rm ba}$=1.256~eV, corresponding to the H2 ZPL at 986~nm. In this case, the extracted coherence decay time is $T_2$=2.63~fs. Although the absolute amplitude of the reconstructed absorbance is approximately a factor of two larger than that obtained from direct photometric measurements (black curve), both spectra match in the near-infrared region. 
For comparison, the dashed curves in Fig.~\ref{fig:NV_2level_model} show reconstructed absorbance spectra assuming that SA arises from  NV$^-$ or NV$^0$ centers, with ZPLs at 637~nm and 575~nm (photon energies of 1.945~eV and 2.155~eV), respectively. In these cases, the extracted coherence decay times are significantly shorter, on the order of 0.15-0.2~fs, and the resulting absorbance spectra differ drastically from those determined by photometric measurements. 
This discrepancy strongly contradicts the assumption that NV centers dominate the absorption saturation at the excitation wavelength of 1032~nm.

Taken together, these results indicate that the observed saturable absorption is most consistently explained by a substantial contribution from H2 defects, whose absorption band partially overlaps with the excitation wavelength. Nevertheless, the experimental response is most likely a sum of contributions from multiple species of defects. Further studies employing tunable femtosecond excitation sources are required to fully unravel the relative roles of individual defects.


\section{Summary and Conclusions}

In this study, nonlinear optical absorption in commercially available quantum-grade NV diamond crystals was investigated using the open-aperture Z-scan technique with femtosecond pulses at a wavelength of $1032~\mathrm{nm}$. A fundamental distinction in the optical response of the investigated materials was observed: while high-purity electronic-grade diamond (EGSC) exhibited standard third-order nonlinear absorption, the nitrogen-doped samples (MCNV and HCNV) displayed a pronounced saturable absorption (SA) effect. This phenomenon intensified with increasing nitrogen and NV concentrations, allowing for a quantitative description of the process using a two-level system model, which yielded a saturation intensity of approximately $I_s \approx 40.0~\mathrm{GW/cm^{2}}$ for the highly doped sample.

Correlating the nonlinear measurements with a detailed analysis of linear transmission spectra allowed for the identification of the microscopic origin of the observed saturation. It was demonstrated that the nonlinear response at $1032~\mathrm{nm}$ is not dominated by NV centers, but rather stems from a substantial contribution of H2 defects (NVN complexes), whose absorption band -- characterized by a zero-phonon line at $986~\mathrm{nm}$ -- partially overlaps with the excitation photon energy. 

The results obtained underscore the critical role of the complex defect landscape in shaping the nonlinear optical properties of diamond, highlighting the necessity of accounting for contributions from ancillary defects, such as H2, in the design of diamond-based nonlinear and quantum photonic devices. The described mechanism plays a crucial role in advanced imaging techniques, particularly in two-photon microscopy of NV-doped diamond structures, where precise control of nonlinear processes is essential.  

\section{Funding}
Narodowe Centrum Nauki (2021/03/Y/ST3/00185, 2023/05/Y/ST3/00135); European Regional Development Fund (POIR.04.02.00-00-D001/20).

\section{Acknowledgments}
 This research was funded in part by
the National Science Centre, Poland, grants 2021/03/Y/ST3/00185 and 2023/05/Y/ST3/00135 within the QuantERA II Programme that has received funding from the European Union’s Horizon 2020 research and innovation programme
under Grant Agreement No 101017733. The study was carried out using research infrastructure purchased with the funds of the European Union in the framework of the Smart Growth Operational Programme, Measure 4.2; Grant No. POIR.04.02.00-00-D001/20, "ATOMIN 2.0 - ATOMic scale science for the INnovative economy"
\section{Disclosures}
The authors declare no conflict of interest.

\section{Data availability}
Data underlying the results presented in this paper are available in Ref. \cite{DATA_2026}

\newcommand{\ZScanOA}{} \label{eq:ZScanOA}
\newcommand{\nielinabsorpcja}{} \label{eq:nielin_absorpcja}
\newcommand{\figNVCAresults}{} \label{fig:NV-CAresults1}

\bibliography{bib}

\end{document}